\documentclass[twocolumn,showpacs,preprintnumbers]{revtex4}

\input{epsf}

\def\CQG{{\it Class. Quantum Gravity} }

\def\PR{{\it Phys. Rev.} }

\def\al{\alpha} \def\be{\beta}  
   
\def\th{\theta}   
\def\la{\lambda}

   \def\Th{\Theta}
   
  \def\mn{{\mu\nu}}

 \def\frac#1#2{{\textstyle{{#1}\over
{#2}}}} 
\def\lsim{\mathrel{\rlap{\lower4pt\hbox{\hskip1pt$\sim$}}
\raise1pt\hbox{$<$}}}
\def\gsim{\mathrel{\rlap{\lower4pt\hbox{\hskip1pt$\sim$}}
\raise1pt\hbox{$>$}}} \def\sqr#1#2{{\vcenter{\vbox{\hrule height.#2pt
\hbox{\vrule width.#2pt height#1pt \kern#1pt \vrule width.#2pt} \hrule
height.#2pt}}}}
\def\square{\mathchoice\sqr66\sqr66\sqr{2.1}3\sqr{1.5}3}
\def\beq{\begin{equation}} \def\eeq{\end{equation}}
\def\beqa{\begin{eqnarray}} \def\eeqa{\end{eqnarray}}

\begin{document}

\title{Nonminimal coupling of perfect fluids to curvature}

\vskip 0.2cm

\author{Orfeu Bertolami\footnote{Also at Instituto de Plasmas e F\'isica Nuclear,
Instituto Superior T\'ecnico.}}
\email{orfeu@cosmos.ist.utl.pt}
\affiliation{Instituto Superior T\'ecnico, Departamento de
F\'{\i}sica, \\Av. Rovisco Pais 1, 1049-001 Lisboa, Portugal}

\author{Francisco S. N. Lobo}
\email{francisco.lobo@port.ac.uk} \affiliation{Institute of
Cosmology \& Gravitation,
             University of Portsmouth, Portsmouth PO1 2EG, UK}
\affiliation{Centro de Astronomia e Astrof\'{\i}sica da
Universidade de Lisboa, Campo Grande, Ed. C8 1749-016 Lisboa,
Portugal}

\author{Jorge P\'aramos\footnotemark[1]}
\email{jorge.paramos@ist.utl.pt}
\affiliation{Instituto Superior T\'ecnico, Departamento de
F\'{\i}sica}

\vskip 0.5cm

\date{\today}

\begin{abstract}

In this work, we consider different forms of relativistic perfect
fluid Lagrangian densities, that yield the same gravitational
field equations in General Relativity (GR). A particularly
intriguing example is the case with couplings of the form
$[1+f_2(R)] {\cal L}_m$, where $R$ is the scalar curvature, which
induces an extra force that depends on the form of the Lagrangian
density. It has been found that, considering the Lagrangian
density ${\cal L}_m=p$, where $p$ is the pressure, the extra-force
vanishes. We argue that this is not the unique choice for the
matter Lagrangian density, and that more natural forms for ${\cal
L}_m$ do not imply the vanishing of the extra-force. Particular
attention is paid to the impact on the classical equivalence
between different Lagrangian descriptions of a perfect fluid.

\vskip 0.5cm

\end{abstract}


\maketitle


\section{Introduction}

Recently, in the context of $f(R)$ modified theories of gravity,
it was shown that a function of $R-$matter coupling induces a
non-vanishing covariant derivative of the energy-momentum,
$\nabla_\mu T^{\mn} \neq 0$. This potentially leads to a deviation
from geodesic motion, and consequently the appearance of an extra
force \cite{Bertolami:2007gv}. Implications, for instance, for
stellar equilibrium in this context have been studied in Ref.
\cite{Bertolami:2007vu}. The equivalence with scalar-tensor
theories with two scalar fields has been considered in Ref.
\cite{Bertolami:2008im}, and a viability stability criterion was
also analyzed in Ref. \cite{Faraoni:2007sn}. This novel coupling
has attracted some attention and, actually, in a recent paper
\cite{Sotiriou:2008it}, this possibility has been applied to
distinct matter contents, such as a massive scalar field and a
dust distribution. Regarding the latter, it was argued that a
``natural choice'' for the matter Lagrangian density for perfect
fluids is ${\cal L}_m=p$, based on Refs.
\cite{Schutz:1970my,Brown:1992kc}, where $p$ is the pressure. This
choice has a particularly interesting application in the analysis
of the $R-$matter coupling for perfect fluids, which implies in
the vanishing of the extra force. However, we point out that,
despite the fact that ${\cal L}_m=p$ does indeed reproduce the
perfect fluid equation of state, it is not unique: Other choices
include, for instance, ${\cal L}_m=-\rho$
\cite{Brown:1992kc,HawkingEllis}, where $\rho$ is the energy
density, or ${\cal L}_m=-na$, where $n$ is the particle number
density, and $a$ is the physical free energy defined as
$a=\rho/n-Ts$, with $T$ being the fluid temperature and $s$ the
entropy per particle. Indeed, all these are on-shell
representations of a more general Lagrangian density, that is,
obtained through back-substitution of the equations of motion into
the related action (see Ref. \cite{Brown:1992kc} for details).
Furthermore, this equivalence is established
within the framework of GR. In this work, we address the issue of
the Lagrangian formulation of perfect fluids in the context of the
proposed model with a non-minimal coupling of the scalar curvature
to matter, as depicted below.

This paper is organized as follows: In Section \ref{Sec:II}, we
review the equations of motion in a curvature-matter coupling; in
Section \ref{Sec:III}, we show the non-uniqueness of the
relativistic perfect matter Lagrangian densities; in Section
\ref{Sec:IV}, we analyze the perfect fluid Lagrangian description
with a non-minimal scalar curvature coupling; and in Section
\ref{Sec:V}, we present our conclusions. Throughout this work, we
use the convention $\kappa=8\pi G=1$ and the metric signature
$(-1,1,1,1)$.

\section{Equation of motion with curvature-matter couplings}
\label{Sec:II}

The action for curvature-matter couplings, in $f(R)$ modified theories
of gravity \cite{Bertolami:2007gv}, takes the following form
\beq \label{model} S=\int \left[{1 \over 2}f_1(R)+\left[1+\lambda
f_2(R)\right]{\cal L}_{m}\right] \sqrt{-g}\;d^{4}x\,, \eeq
\noindent where $f_i(R)$ (with $i=1,2$) are arbitrary functions of
the curvature scalar $R$ and ${\cal L}_{m}$ is the Lagrangian
density corresponding to matter.

Varying the action with respect to the metric $g_{\mu \nu }$
yields the field equations, given by
\beqa \nonumber
&& F_1R_{\mu \nu }-{1\over 2}f_1g_{\mu \nu }-\nabla_\mu \nabla_\nu
F_1+g_{\mu\nu}\square F_1= (1+\lambda f_2)T_{\mu
\nu }\\ && -2\lambda F_2{\cal L}_m R_{\mu\nu}
   +2\lambda(\nabla_\mu
\nabla_\nu-g_{\mu\nu}\square){\cal L}_m F_2 \,, \label{field}
\eeqa
\noindent where we have denoted $F_i(R)=f'_i(R)$, and the prime
represents the derivative with respect to the scalar curvature.
The matter energy-momentum tensor is defined as
\beq T_{\mu \nu}=-{2 \over \sqrt{-g}}{\delta(\sqrt{-g}{\cal
L}_m)\over \delta(g^{\mu\nu})} \,. \label{defSET} \eeq
Now, taking into account the generalized Bianchi identities, one
deduces the following corrected conservation equation
\beq \nabla^\mu T_{\mu \nu }={\lambda F_2\over 1+\lambda
f_2}\left[g_{\mu\nu}{\cal L}_m- T_{\mu \nu}\right]\nabla^\mu R\,.
\label{cons1} \eeq
\noindent If one considers the equivalence with a scalar field
theory (with two scalar fields, $\phi = R$ and $\psi = {\cal
L}_m$) \cite{Bertolami:2008im}, it is clear that the non-minimal
coupling between curvature and matter yields an exchange of energy
and momentum between the latter and these scalar fields.

In the following, consider the equation of state for a perfect
fluid
\beq T_{\mu \nu }=\left( \rho +p\right) U_{\mu }U_{\nu }+pg_{\mu
\nu }\,, \eeq
\noindent where $\rho$ is the energy density and $p$, the
pressure, respectively. The four-velocity, $U_{\mu }$, satisfies
the conditions $U_{\mu }U^{\mu }=-1$ and $U^{\mu }U_{\mu ;\nu
}=0$.

Introducing the projection operator $h_{\mu \nu}=g_{\mu\nu}+U_{\mu
}U_{\nu}$, gives rise to non-geodesic motion governed by the
following equation of motion for a fluid element
\beq {dU^{\mu}\over ds}+\Gamma _{\alpha
\beta}^{\mu}U^{\alpha}U^{\beta} =f^{\mu}\,, \label{eq1} \eeq
\noindent where the extra force, $f^{\mu}$, is given by
\beq f^{\mu}={1\over \rho +p}\left[{\lambda F_2\over 1+\lambda
f_2}\left({\cal L}_m-p\right)\nabla_\nu R+\nabla_\nu p \right]
h^{\mu\nu }\,.
     \label{force}
\eeq
\noindent One verifies that the first term vanishes for the
specific choice of ${\cal L}_m=p$, as noted by Ref.
\cite{Sotiriou:2008it}. However, as emphasized in the
Introduction, this is not the unique choice for the Lagrangian
density of a perfect fluid, as we shall outline below.

\section{Relativistic perfect fluid matter Lagrangian densities}
\label{Sec:III}

In this section, we follow Ref. \cite{Brown:1992kc} closely, and
review the Lagrangian formulation of a perfect fluid in the
context of GR. The action is presented in terms of
Lagrange multipliers along the Lagrange coordinates $\alpha^A$ in
order to enforce specific constraints, and is given by
\beq
  S_m =\int d^4x \left[-\sqrt{-g} \;\rho(n,s)+J^\mu\left(\varphi_{,\mu}
  +s\theta_{,\mu}+\beta_A \alpha^{A}_{,\mu}\right)\right] ~.
  \label{fluid-action}
\eeq
\noindent Note that the action $S_m
=S(g_{\mu\nu},J^\mu,\varphi,\theta,s,\alpha^A,\beta_A)$ is a
functional of the spacetime metric $g_{\mu\nu}$, the entropy per
particle $s$, the Lagrangian coordinates $\alpha^A$, and spacetime
scalars denoted by $\varphi$, $\theta$, and $\beta_A$, where the
index $A$ takes the values 1, 2, 3 (see Ref. \cite{Brown:1992kc}
for details). The physical interpretation of these parameters is
given below.

The vector density $J^\mu$ is interpreted as the flux vector of
the particle number density, and defined as
$J^\mu=\sqrt{-g}\,nU^\mu$. The particle number density is given by
$n=|J|/\sqrt{-g}$, so that the energy density is given as a
function $\rho=\rho(|J|/\sqrt{-g},s)$.

Varying the action with respect to the metric, and using the
definition given by Eq. (\ref{defSET}), provides the stress-energy
tensor for a perfect fluid
\beq T^{\mu\nu}=\rho \, U^\mu U^\nu +\left(n{\partial \rho\over
\partial n}-\rho\right)\left(g^{\mu\nu}+U^\mu U^\nu \right)\,,
\eeq \noindent with the pressure defined as \beq \label{pressure
definition} p=n{\partial \rho\over \partial n}-\rho\,. \eeq
\noindent Note that this definition of pressure is in agreement
with the First Law of Thermodynamics, $d\rho=\mu\, dn + nT ds$.
The latter shows that the equation of state can be specified by
giving the function $\rho(n,s)$, {\it i.e.}, the energy density as
a function of number density and entropy per particle. The
quantity $\mu=\partial \rho/\partial n=(\rho+p)/n$ is defined as
the chemical potential, which is the energy per particle required
to inject a small amount of fluid into a fluid sample, maintaining
a constant sample volume and a constant entropy per particle $s$.
In addition, when imposing the stress-energy tensor covariant
conservation, {\it i.e.}, $T^{\mu\nu}_{;\nu}=0$, the perfect fluid
also implies the covariant conservation of particle number, given
by $(nU^\mu)_{;\mu}=0$.

The variation of the action with respect to $J^\mu$, $\varphi$,
$\theta$, $s$, $\alpha^A$ and $\beta_A$, provides the following
equations of motion,
\beqa {\delta S\over \delta J^\mu}&=&\mu U_\mu +\varphi_{,\mu}
+s\theta_{,\mu}+\beta_A\alpha^{A}_{,\mu}=0\,,
  \label{eqsmotion1} \\
{\delta S\over \delta \varphi}&=&-J^{\mu}_{,\mu}=0 \,,
  \label{eqsmotion2} \\
{\delta S\over \delta \theta}&=&-(sJ^{\mu})_{,\mu}=0\,,
  \label{eqsmotion3} \\
{\delta S \over \delta s}&=&-\sqrt{-g}{\partial \rho\over \partial
s} +\theta_{,\mu}J^\mu=0 \,,
  \label{eqsmotion4} \\
{\delta S\over \delta \alpha^A}&=&-(\beta_A J^\mu)_{,\mu}=0 \,,
  \label{eqsmotion5} \\
{\delta S\over \delta \beta_A}&=&\alpha^{A}_{,\mu} J^\mu=0
\label{eqsmotion6} \,. \eeqa
\noindent The first relationship, Eq. (\ref{eqsmotion1}), provides
the velocity-representation of the $4-$velocity; the second
equation, Eq. (\ref{eqsmotion2}), reflects the particle number
conservation, {\it i.e.}, $(nU^\mu)_{;\mu}={1\over
\sqrt{-g}}J^{\mu}_{,\mu}=0$; Eq. (\ref{eqsmotion3}) translates the
entropy exchange constraint; Eq. (\ref{eqsmotion4}) provides the
identification of $T=\theta_{,\mu}U^\mu={1\over n}{\partial
\rho\over \partial s}|_n$ after comparing it with the First Law of
Thermodynamics; Eq. (\ref{eqsmotion5}) reflects the constancy of
the parameter $\beta_A$ along the fluid flow lines; and finally,
Eq. (\ref{eqsmotion6}) restricts the fluid $4-$velocity to be
directed along the flow lines of constant $\alpha^{A}$.

One may now infer the physical interpretation for the respective
parameters. The scalar field $\varphi$ is interpreted as a
potential for the chemical free energy $f$, and is a Lagrange
multiplier for $J^{\mu}_{,\mu}$, the particle number conservation.
The scalar fields $\beta_A$ are interpreted as the Lagrange
multipliers for $\alpha^{A}_{,\mu} J^\mu=0$, restricting the fluid
$4-$velocity to be directed along the flow lines of constant
$\alpha^{A}$.

Note that taking into account Eq. (\ref{eqsmotion1}), and the
definitions $J^\mu=\sqrt{-g}\,nU^\mu$ and $\mu=(\rho+p)/n$, the
action Eq. (\ref{fluid-action}) reduces to the on-shell Lagrangian
density ${\cal L}_{m(1)} = p$, with the action given by \beq
  S_m =\int d^4x \sqrt{-g}\, p \,,
  \label{fluid-actionP}
\eeq
\noindent which is the form considered in Ref.
\cite{Schutz:1970my}. One should bear in mind that this on-shell
Lagrangian density yields the equations of motion
(\ref{eqsmotion1})-(\ref{eqsmotion6}) only if one considers that
the pressure is functionally dependent on the previously
considered fields $\varphi$, $s$, $\th$, $\be_A$, $\al^A$, and on
the current density $J^\mu$.

Now, it was a Lagrangian density given by ${\cal L}_m=p$ that the
authors of Ref. \cite{Sotiriou:2008it} use to obtain a vanishing
extra-force due to the non-trivial coupling of matter to the
scalar curvature $R$. For concreteness, replacing ${\cal L}_m=p$
in Eq. (\ref{force}), one arrives at the general relativistic
expression
\beq f^{\mu}={h^{\mu\nu}\nabla_\nu p \over \rho+p}  ~. \eeq
\noindent However, the on-shell degeneracy of the Lagrangian
densities arises from adding up surface integrals to the action.
For instance, consider the following surface integrals added to
the action Eq. (\ref{fluid-action}),
\beqa &-\int d^4x (\varphi J^\mu)_{,\mu} ~,~~~  -\int d^4x (\theta
sJ^\mu)_{,\mu}\,,~~
     \nonumber   \\
& -\int d^4x (J^\mu \beta_A\alpha^A)_{,\mu}  ~,
  \nonumber
\eeqa
\noindent so that the resulting action takes the form
\beqa
  S&=&\int d^4x \Big[-\sqrt{-g}\, \rho(n,s)-\varphi J^{\mu}_{,\mu}
   \nonumber    \\
  &-&\theta(sJ^{\mu})_{,\mu}-\alpha^{A}(\beta_AJ^{\mu})_{,\mu}\Big] \,.
  \label{fluid-action2}
\eeqa \noindent Note that this action reproduces the equations of
motion, Eqs. (\ref{eqsmotion1})-(\ref{eqsmotion6}). Taking into
account the latter, the action reduces to \beq
  S_m =-\int d^4x \sqrt{-g}\, \rho \,,
  \label{fluid-action-rho}
\eeq \noindent {\it i.e.}, the on-shell matter Lagrangian density
takes the following form ${\cal L}_m=-\rho$; as before, $\rho$ is
dependent on the original fields present in the action Eq.
(\ref{fluid-action}). This choice is also considered for
isentropic fluids, where the entropy per particle is constant
$s={\rm const}$ \cite{Brown:1992kc,HawkingEllis}. For the latter,
the First Law of Thermodynamics indicates that isentropic fluids
are described by an equation of state of the form
$a(n,T)=\rho(n)/n-sT$ \cite{Brown:1992kc} (see Ref.
\cite{Bertolami:2008pa} for a bulk-brane discussion of this
choice).

For this specific choice of ${\cal L}_{m(2)}=-\rho$ the extra
force takes the following form: \beq f^{\mu}=\left(-{\lambda
F_2\over 1+\lambda f_2}\nabla_\nu R+{1\over \rho +p}\nabla_\nu p
\right)h^{\mu\nu}\,.
    \label{force2}
\eeq \noindent An interesting characteristic is that the term
related to the specific curvature-matter coupling is independent on the
energy-matter distribution.

Another interesting action functional is given by the equation of
state of the physical free energy as a function of the number
density and the temperature, $a(n,T)$. For this we follow the
reasoning of Ref. \cite{Brown:1992kc}. For instance,
solving Eq. (\ref{eqsmotion4}) for $s$ as a function of $n$ and
$T$ (using the definition $T=\theta_{,\mu}J^\mu/|J|$), and finally
eliminating $s$ from the action Eq. (\ref{fluid-action}), yields
\beq\label{an}
S_m =\int d^4x \left[-|J|\;a(n,T)+J^\mu\left(\varphi_{,\mu}
  +\beta_A \alpha^{A}_{,\mu}\right)\right] \,.
\eeq
\noindent Using the definitions $n=|J|/\sqrt{-g}$ and
$T=\theta_{,\mu}J^\mu/|J|$, and varying the action with respect to
$J^\mu$, one ends up with the equation of motion Eq.
(\ref{eqsmotion1}). The remaining equations of motion are readily
obtained by varying $S$ with respect to $\varphi$, $\theta^A$ and
$\beta_A$. It is simple to show that this action also provides the
perfect fluid stress-energy tensor. As before, one may consider
the addition of the following surface integrals to Eq.
(\ref{fluid-action})
\beqa -\int d^4x (\varphi J^\mu)_{,\mu} \,,
   \qquad
   -\int d^4x (J^\mu \beta_A\alpha^A)_{,\mu}  \,,
  \nonumber
\eeqa \noindent so that the action takes the following form
\beq\label{an2} S_m =\int d^4x \left(-\sqrt{-g}\;na\right) \,.
\eeq
\noindent The matter Lagrangian density is given by ${\cal
L}_{m(3)}=-na$. The extra force in terms of this Lagrangian
density yields the following expression:
\beq
f^{\mu}={1 \over \rho +p}\left[-{\lambda F_2 \over 1+\lambda
f_2}(na+p)\nabla_\nu R+\nabla_\nu p \right] h^{\mu\nu} ~.
     \label{force3}
\eeq

Hence, it is clear that no immediate conclusion may be extracted
regarding the additional force imposed by the non-minimal coupling
of curvature to matter, given the different available choices for
the Lagrangian density; moreover, one could doubt the validity of
a conclusion that allows for different physical predictions
arising from these apparently equivalent Lagrangian densities.

Thus, from the above point of view, there is no particular reason
to regard the choice of the on-shell Lagrangian density ${\cal
L}_{m(1)} = p$ as preferable over the others we have discussed
above. However, this degeneracy of the Lagrangian density of a
perfect fluid, which does not appear in GR, is
rather intriguing and object of further discussion in the next
section.

\section{Perfect fluid Lagrangian description with non-minimal
scalar curvature coupling}\label{Sec:IV}

There is a caveat in above treatment, which can easily pass
ignored: The discussion of the Lagrangian density-dependence of
the extra force given by Eq. (\ref{force}), and degeneracy
thereof, implicitly admits that the equivalence between different
on-shell Lagrangian densities holds. However, the latter is
established in GR, and may not be valid in the more general model
considered here.

Clearly, one may argue that two Lagrangian densities are
equivalent if both generate the same energy-momentum tensor, and
if variation of the corresponding actions yields the same
equations of motion (\ref{eqsmotion1})-(\ref{eqsmotion6}). As has
been shown above, the several on-shell Lagrangian densities ${\cal
L}_{m(1)} = p$, ${\cal L}_{m(2)} = -\rho$, ${\cal L}_{m(3)} = -na$
are all equivalent to the original, ``bare'' Lagrangian density,
\beq {\cal L}_m = -\rho(n,s)+{J^\mu \over
\sqrt{-g}}\left(\varphi_{,\mu}
  +s\theta_{,\mu}+\beta_A \alpha^{A}_{,\mu}\right)\,. \label{original}
  \eeq
\noindent Hence, one must attempt to retrace the derivation of the
classical equivalence leading to these on-shell quantities.
Clearly, if one simply includes the $\left[1 + \la f_2(R) \right]$
factor of Eq. (\ref{model}) into action Eq. (\ref{fluid-action}),
that is,
\beqa
  S&=&\int d^4x \left[1 + \la f_2(R) \right] \times
  \nonumber \\
  && \left[-\sqrt{-g} \;\rho(n,s)+J^\mu\left(\varphi_{,\mu}
  +s\theta_{,\mu}+\beta_A \alpha^{A}_{,\mu}\right)\right] ,
\eeqa
\noindent then the equations of motion
(\ref{eqsmotion1})-(\ref{eqsmotion6}) are unaffected, as variation
with respect to each field yields only a global factor $\left[1 +
\la f_2(R) \right]$.

However, the guiding principle behind the proposal first put
forward in Ref. \cite{Bertolami:2007gv} is to allow for a
non-minimal coupling between curvature and matter. Thus, the
modification of the perfect fluid action Eq. (\ref{fluid-action})
should only affect the terms that show a minimal coupling between
curvature and matter, {\it i.e.}, those multiplied by $\sqrt{-g}$.
For this reason, the current density term, which is not coupled to
curvature, should not be altered. This yields
\beqa
  S'_m&=&\int d^4x \bigg{[}- \sqrt{-g} \left[1 + \la f_2(R) \right]
  \rho(n,s)
  \nonumber \\
  && +J^\mu\left(\varphi_{,\mu} +s\theta_{,\mu}
  +\beta_A \alpha^{A}_{,\mu}\right)\bigg{]} \,,
  \label{modified fluid-action}
\eeqa
\noindent The equations of motion (\ref{eqsmotion2}),
(\ref{eqsmotion3}), (\ref{eqsmotion5}) and (\ref{eqsmotion6}) are
unchanged, while Eqs. (\ref{eqsmotion1}) and (\ref{eqsmotion4})
read
\beqa {\delta S\over \delta J^\mu}&=&\mu \left[1 + f_2(R) \right]
U_\mu +\varphi_{,\mu} +s\theta_{,\mu}+\beta_A\alpha^{A}_{,\mu}=0
\,, \nonumber
  \label{eqsmotion1a} \\
{\delta S \over \delta s}&=&-\sqrt{-g} \left[1 + f_2(R) \right]
{\partial \rho\over \partial s} +\theta_{,\mu}J^\mu=0 \,.
  \label{eqsmotion4a}
\eeqa
\noindent This results from the coupling of the
variables $J^\mu$ and $s$ with the factor $\left[ 1+f_2(R)
\right]$ (since $n = |J| / \sqrt{-g}$). Recalling that $J^\mu =
\sqrt{-g} n U^\mu$, one obtains
\beqa  &&-\left[ 1 + \la f_2(R) \right]\mu U_\mu = \varphi_{,\mu}
+s\theta_{,\mu}+\beta_A\alpha^{A}_{,\mu} \,, \\
&&T = {1 \over n}{\partial \rho \over \partial s}\Bigg|_n = {1
\over 1 + \la f_2(R)} \th_{,\mu}U^\mu \,, \eeqa
\noindent so that both the velocity representation and the
temperature reflect the non-minimal coupling of curvature to
matter.

One may now proceed and substitute the modified equations of
motion into action (\ref{modified fluid-action}), in order to
obtain the new on-shell Lagrangian density,
\beq
  S'_m=\int d^4x \sqrt{-g} \left[1 + \la f_2(R) \right] p \,.
  \label{modified on-shell}
\eeq
\noindent Hence, one concludes that the on-shell Lagrangian
density ${\cal L}_{m(1)} = p$ is also obtained in the considered
scenario. By including extra surface integrals, a similar
procedure (not pursued here) also yields the previously discussed
forms ${\cal L}_{m(2)} = -\rho$, ${\cal L}_{m(3)} = -na$.

\subsection{Gravitational field equations and the nonequivalence
between on-shell and bare Lagrangian densities}

The above discussion confirms that one may adopt any particular
on-shell Lagrangian density as a suitable functional for
describing a perfect fluid, therefore leading to the issue of
distinguishing between different predictions for the extra force.
However, this is not quite correct: although the above Lagrangian
densities ${\cal L}_{m(i)}$ are indeed obtainable from the
original action, it turns out that they are not equivalent to the
original Lagrangian density ${\cal L}_{m}$. Indeed, this
equivalence demands that not only the equations of motion of the
fields describing the perfect fluid remain invariant, but also
that the gravitational field equations do not change.

Recall that the terms in the field equations (\ref{field}) which
depend on ${\cal L}_m$ arise from the presence of the non-minimal
coupling $[1 + \la f_2(R)]$. However, the formulation of a perfect
fluid action functional includes the presence of a current density
term, plus eventual surface integral terms $B^\mu_{;\mu}$. Writing
${\cal L}_c = -\rho(n,s)$, $V_\mu \equiv
\varphi_{,\mu}+s\theta_{,\mu}+\beta_A\alpha^{A}_{,\mu}$, for
simplicity, then
\beq
  S'_m =\int d^4x \left[\sqrt{-g}\left[ 1 + \la f_2(R) \right]
  {\cal L}_c+J^\mu V_\mu + B^\mu_{;\mu} \right] \,,
\eeq
\noindent one can see that only the non-minimal coupled term
${\cal L}_c$ appears in the field equations, as variations with
respect to $g^\mn$ of the remaining terms vanish:
\beqa \nonumber
&& F_1R_{\mu \nu }-{1\over 2}f_1g_{\mu \nu }-\nabla_\mu \nabla_\nu
F_1+g_{\mu\nu}\square F_1= (1+\lambda f_2)T_{\mu
\nu }\\ && -2\lambda F_2{\cal L}_c R_{\mu\nu}
   +2\lambda(\nabla_\mu
\nabla_\nu-g_{\mu\nu}\square){\cal L}_c F_2 \,, \label{field2}
\eeqa
\noindent Clearly, the appropriate energy-momentum tensor is still
obtained from ${\cal L}_c$, definitions (\ref{pressure
definition}) and relations $U^\mu U_\mu = -1$ and $J^\mu
=\sqrt{-g} n U^\mu$.

One arrives not at a paradox, but a tautology: different
predictions for non-geodesic motion result from different forms of
the gravitational field equations. Therefore, the equivalence
between different on-shell Lagrangian densities ${\cal L}_{m(i)}$
and the original quantity ${\cal L}_{m}$ is broken, so that one
can no longer freely choose between the available forms.

By the same token, the additional extra force is unique, and
obtained by replacing ${\cal L}_c = -\rho$ into Eq. (\ref{force}),
yielding expression (\ref{force2}), here repeated for convenience
\beq f^{\mu}=\left(-{\lambda F_2\over 1+\lambda f_2}\nabla_\nu
R+{1\over \rho +p}\nabla_\nu p \right)h^{\mu\nu}\,. \eeq

\subsection{Null dust case}

Following Ref. \cite{Sotiriou:2008it}, it is interesting to
analyze the generalized conservation law, given by Eq.
(\ref{cons1}), in the case of a null dust matter distribution. The
latter is defined as the particular case of a perfect fluid with
vanishing pressure, $p = 0$. This is usually interpreted as
expressing weakly interacting non-relativistic particles, with
$\rho c^2 \gg p \approx 0$. However, given the previous discussion
of the functional description of a perfect fluid, where the
pressure is not an independent quantity, but defined by Eq.
(\ref{pressure definition}), a more rigorous (and physically
compatible) formulation corresponds to an isentropic ($s = {\rm
const}$) perfect fluid with an equation of state of the form
$\rho(n)=n\mu$, with a constant chemical potential $\mu$.

The authors of Ref. \cite{Sotiriou:2008it} conclude that the extra
force arising due to the non-minimal coupling of dust with the
scalar curvature does not lead to non-geodesic motion, as it
preserves parallel transport (and only changes the
parameterization of the geodesic). However, this result arises
from the particular choice ${\cal L}_m = p$, which is commonly
used in the framework of GR. As the previous discussion has shown,
in the context of the considered curvature-matter coupling model, one
cannot freely chose between available on-shell Lagrangian
densities, since these do not lead to the same gravitational field
equations \footnote{Even if one could opt for equivalent on-shell
Lagrangian densities in the context of the model considered, the
choice ${\cal L}_m = p$ for null dust is troublesome, since then
this quantity vanishes by definition.}.

Instead, inserting the component of the original Lagrangian
density that couples to the geometry, ${\cal L}_c = -\rho $ into
Eq. (\ref{cons1}), and considering the energy-momentum tensor
$T_\mn = \rho U_\mu U_\nu$, one arrives at the following
relationship
\beqa
&& \left( U^\mu \nabla_\mu U_\nu + U_\nu \nabla^\mu U_\mu +
U_\nu U_\mu \nabla^\mu \right) \rho =
\nonumber \\
&& -{\la F_2 \rho \over 1 + \la f_2 } \left(g_\mn + U_\mu U_\nu
\right) \nabla^\mu R \,. \eeqa
\noindent Following the notation of Ref. \cite{Sotiriou:2008it},
one writes
\beq \Th = \nabla^\mu U_\mu + {1 \over \rho}U_\mu \nabla^\mu \rho
+ {\la F_2 \over 1 + \la f_2} U_\mu \nabla^\mu R\,, \eeq
\noindent obtaining
\beq U^\mu \nabla_\mu U_\nu = -\Th U_\nu - {\la F_2 \over 1 + \la
f_2 } \nabla_\nu R\,,   \label{non-geodesic}\eeq
\noindent which clearly shows that parallel transport is no longer
conserved, and one concludes that non-geodesic motion is also
followed by pressureless dust.

\section{Conclusions}\label{Sec:V}

In this work we have discussed the degeneracy of Lagrangian
densities for a perfect fluid, in the context of a gravity model
where matter is coupled non-minimally with the scalar curvature.
This degeneracy problem is well known in the context of GR, but in
the discussed non-minimally coupled model possesses some new and
rather surprising features, such as non-geodesical motion (first
discussed in Ref. \cite{Bertolami:2007gv}; see Ref.
\cite{Lobo:2008sg} for a recent review); this dependency on the
choice of the Lagrangian density was pointed out in Ref.
\cite{Sotiriou:2008it}.

We show that this degeneracy does not appear in the considered
model, since different on-shell Lagrangian densities which are
classically equivalent do not yield the same gravitational field
equations. Instead, we conclude that only the part of the original
Lagrangian density ${\cal L}_m$ that is coupled to the geometry
(via the $\sqrt{-g}$ factor) appears in these field equations.
Hence, it follows that the motion of test particles is necessarily
non-geodesic, if the non-minimal nature of the coupling between
matter and curvature is properly accounted in the onset of fluid
treatment.

However, we should point out that this study only solves the issue
of an apparent degeneracy due to the classical equivalence between
on-shell Lagrangian densities. This should not be seen as an
exhaustive account of the overall problem, since it only lifts
this degeneracy for a particular original Lagrangian density
${\cal L}_m$. One can take a Lagrangian density different from
that of Eq. (\ref{fluid-action}) to begin with, which also gives a
full account of the behavior of a perfect fluid, describing both
the correct energy-momentum tensor as well as its thermodynamics.
As an example, if  the bare Lagrangian density of Ref. \cite{Taub}
is considered (see Ref. \cite{Poplawski:2008as} for a discussion),
one obtains an extra force that is corrected by a factor linearly
dependent on the Helmholtz free energy.

If one takes an initial ${\cal L}_m$ that is functionally
different from the one adopted in this study, and that still
enables a convenient description of a perfect fluid (and is
suitably interpreted through the use of the First Law of
Thermodynamics), then one could obtain different results for the
predicted extra force. This might lead to two different
interpretations: One can conjecture that there must exist an
underlying principle or symmetry that yields a unique Lagrangian
density for a perfect fluid, so that different extra force
predictions stem from an incomplete action description of it;
however, one might also posit that different extra forces arising
from different Lagrangian densities are physically
distinguishable. If so, the model under scrutiny would serve to
discriminate between different fluids that share the same
energy-momentum tensor (and are thus ``perfect''), but have
different thermodynamic formulations. In the authors' opinion,
this has not been given due attention in the literature, most
likely because arbitrary gravitational field equations depending
on the matter Lagrangian have not often been the object of
scrutiny.

In fact, to judge which matter Lagrangian density is the ``natural'' one
depends, to
some extent, on the independent variables that are considered.
In this respect, an
interesting avenue for future research would be to consider the
non-minimal curvature-matter coupling using
velocity-potentials \cite{Schutz:1970my}. In this case, there are no
constraints in the action principle, and one could compare this analysis
with the one where the comoving Lagrangian coordinates label the fluid
elements and that exhibits constraints \cite{Lusanna:2000mu} (we
refer the reader to Ref. \cite{Schutz:1977df} for a proof of the
need for constraints and related issues). However,
this is not a trivial task, as no Lagrangian is unique, even in the
presence of the non-minimal coupling, since it is invariant under
the addition of a divergence, as mentioned above. Work along these
lines is presently underway.

\begin{acknowledgments}

The authors would like to thank David Brown, Luca Lusanna and
Bernard Schutz for helpful comments and Thomas Sotiriou for his comments
on an earlier
version of this manuscript. O.B. acknowledges the partial
support of the Funda\c{c}\~{a}o para a Ci\^{e}ncia e a Tecnologia
(FCT) project $POCI/FIS/56093/2004$. F.S.N.L. was funded by FCT
through the grant $SFRH/BPD/26269/2006$. The work of J.P. is
sponsored by FCT through the grant $SFRH/BPD/23287/2005$.

\end{acknowledgments}

\end{document}